\documentstyle[preprint,prc,aps,epsfig]{revtex}
\tightenlines

\begin{document}

\title{Low-spin spectroscopy of $^{50}$Mn}
 
\author{ P.~von Brentano$\,^1$, A.~Dewald$\,^1$, C.~Frie\ss{}ner$\,^1$, 
         A.~F.~Lisetskiy$\,^1$, N.~Pietralla$\,^{1,2}$, 
         I.~Schneider$\,^1$, A.~Schmidt$\,^1$,
         T.~Sebe$\,^3$, T.~Otsuka$\,^3$ \\}

\address{$^1\,$ Institut f\"ur Kernphysik, Universit\"at zu K\"oln, 
                50937 K\"oln, Germany\\ E-mail: brentano@ikp.uni-koeln.de \\ 
         $^2\,$ Wright Nuclear Structure Laboratory, Yale University, 
                New Haven, \\ Connecticut 06520-8124, USA \\ 
                E-mail: pietrall@galileo.physics.yale.edu \\
         $^3\,$ Department of Physics, University of Tokyo, Hongo,
                Bunkyo-ku, \\ Tokyo 113-0033, Japan \\
                E-mail: otsuka@phys.s.u-tokyo.ac.jp}

\maketitle

\begin{abstract}
 The data on low spin states in the odd-odd nucleus $^{50}$Mn 
  investigated  with the  $^{50}$Cr\,(p,n$\gamma$)$^{50}$Mn fusion 
  evaporation reaction at the FN-{\sc Tandem} accelerator in Cologne are 
  reported. Shell model and collective rotational model interpretations of 
  the data are given.
\end{abstract}


\section{Introduction}

Atomic nuclei along N=Z line where the relative proximity of the 
neutron and proton Fermi surfaces favors the proton-neutron (pn) correlations 
on the equal footing with neutron-neutron (nn) and proton-proton (pp) ones 
are proved to give a unique opportunity to study the different isospin 
modes of pn interaction in various contexts 
\cite{Schneider99,Fries98,Schmidt,Lis99,tokyo,Sve98,Sko98,Rudolph,Lenzi99,OLeary99,Vin98,Faes99,Isa97,Sat97,Buc97,Ots98,Cast91,Lang96,Mac,Vogel,Mac2,Poves}. 

In contrast to pp (T=1,T$_z$=-1) and nn (T=1,T$_z$=1) configurations the pn 
pairs can either form isovector (T=1,T$_z$=0) or isoscalar (T=0,T$_z$=0) 
states. Isovector pn correlations manifest themselves in a similar fashion 
to like-nucleon correlations that is evident from the properties of the T=1 
isobaric analogue nuclei while the pn interaction in isoscalar 
channel (T=0,T$_z$=0) is much less understood and is currently the subject 
of active debate \cite{Lang96,Mac,Vogel,Mac2}.  
In the light of this interest 
recently the low-spin structure of odd-odd  N=Z nuclei $^{54}$Co 
\cite{Schneider99},  $^{46}$V \cite{Fries98} and  $^{50}$Mn \cite{Schmidt} 
was studied in Cologne. The most interesting findings  of our studies are 
strong isovector M1 transitions which have simple explanation in terms of two 
particle quasideuteron configurations \cite{Lis99} and  collective 
rotational properties of low-lying states in $^{46}$V \cite{tokyo} and 
$^{50}$Mn.  

This contribution present new results for $^{50}$Mn establishing the 
equivalence of full pf-shell model and deformed  mean field treatments 
of the low-energy structures in $^{50}$Mn.  

\section{Collective and quasideuteron properties of yrast states in $^{50}$Mn}

\begin{figure*}[htb]
\epsfxsize 15.0cm \centerline{\epsfbox{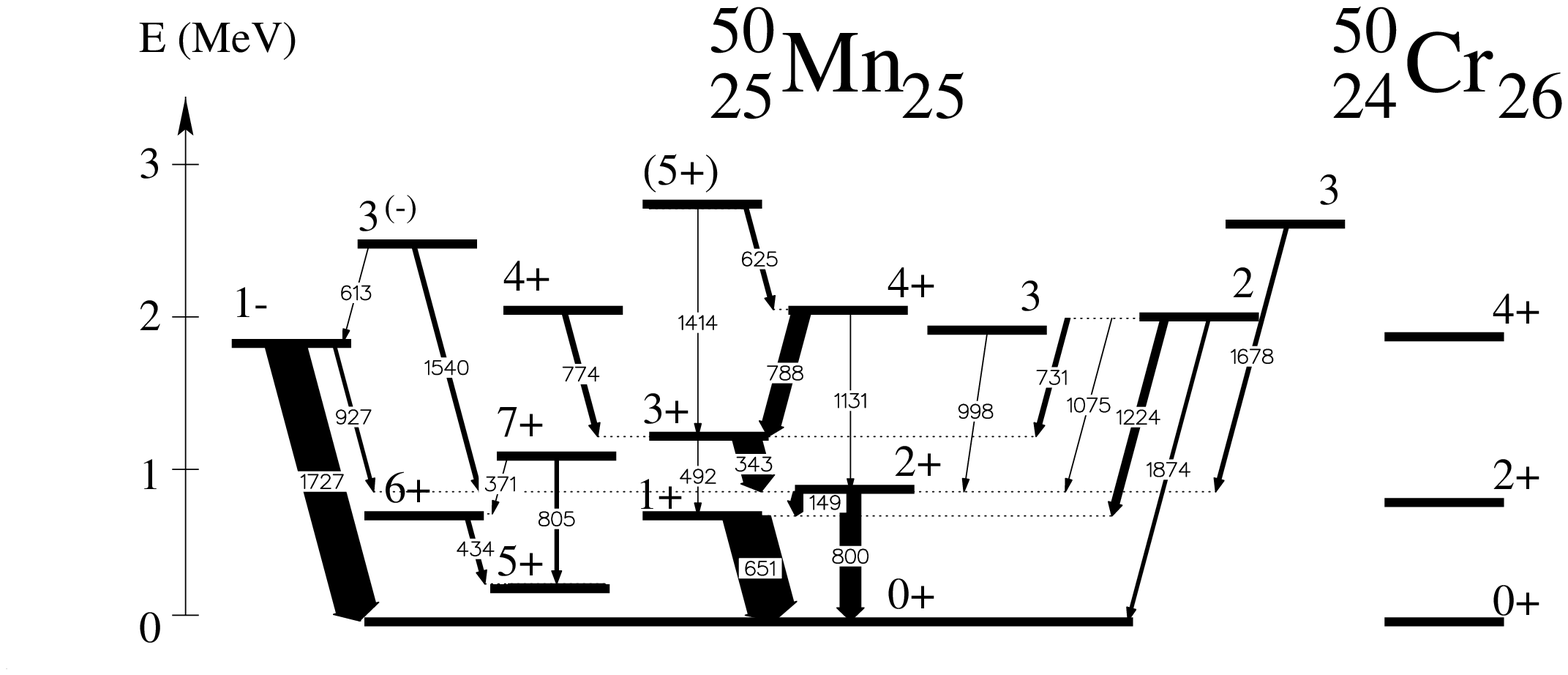}}
\caption{\label{cutscheme}Part of the level scheme of $^{50}$Mn, 
  including only those levels
  for which definite spin or parity quantum numbers are known. The
  width of the arrows corresponds to the relative intensity of the
  $\gamma$-transitions observed in the present reaction. In the right 
panel of the figure the low-lying T=1 states of $^{50}$Cr are shown.
Adapted from A.~Schmidt, {\em et. al,} \protect\cite{Schmidt}.}
\end{figure*}

Recently we have investigated the low-spin structure of the 
odd-odd $N=Z$ nucleus $^{50}$Mn up to an excitation energy of 
$E_x \approx 3.6$~MeV \cite{Schmidt}.
Low spin states of $^{50}$Mn were populated using the fusion
evaporation reaction $^{50}$Cr\,(p,\,n$\gamma$)$^{50}$Mn at a 
proton beam energy $E_p = 15$~MeV. 
The beam was delivered by the FN-{\sc Tandem} accelerator of 
the University of Cologne. A part of the low spin level scheme of 
$^{50}$Mn, which could be 
determined from our $\gamma\gamma$-coincidence data, is shown in 
Fig.\,\ref{cutscheme}.
From the analysis of coincidence spectra, 25 new transitions were
placed in the level scheme, establishing 16 new levels. 
In total, we could assign new spin quantum numbers to six levels. 
Nine new intensity ratios and eight new multipole mixing ratios $\delta$ 
were determined. Our new data together with some 
recent high spin data for $^{50}$Mn from C.~E. Svensson, {\em et al.} 
\cite{Sve98} give a consistent and extensive level scheme for $^{50}$Mn.

The $0^+_1$ ground state, the $2^+_1$ state at 800~keV and  
the $4^+$ states at 1931~keV ( not at 1917 keV as it was reported previously 
by C.~E. Svensson, {\em et al.} \cite{Sve98}) in $^{50}$Mn are 
interpreted as the $T=1$ isobaric analogues of the $0^+_1$ 
ground state, the $2^+_1$ state at 783~keV and the $4^+_1$ state 
at 1882~keV respectively in the isobaric nucleus $^{50}$Cr ( see Fig.\,
\ref{cutscheme}). The interpretation was done using new 
measured intensity and multipole mixing ratios for the $^{50}$Mn. 

The experimental data were compared to shell model (SM)
calculations of the positive parity states of $^{50}$Mn in the full
pf-shell configurational space without truncation. Two different 
nucleon-nucleon residual interactions were considered in \cite{Schmidt}: 
the KB3 interaction, 
adopted from Ref.\,\cite{KB3} and the FPD6 interaction taken from 
Ref.\,\cite{fpd6}. Close agreement with experiment was obtained for 
observables (level scheme, intensity and multipole mixing ratios) near the 
ground state. 

\begin{figure*}[htb]
\epsfxsize 16.8cm \centerline{\epsfbox{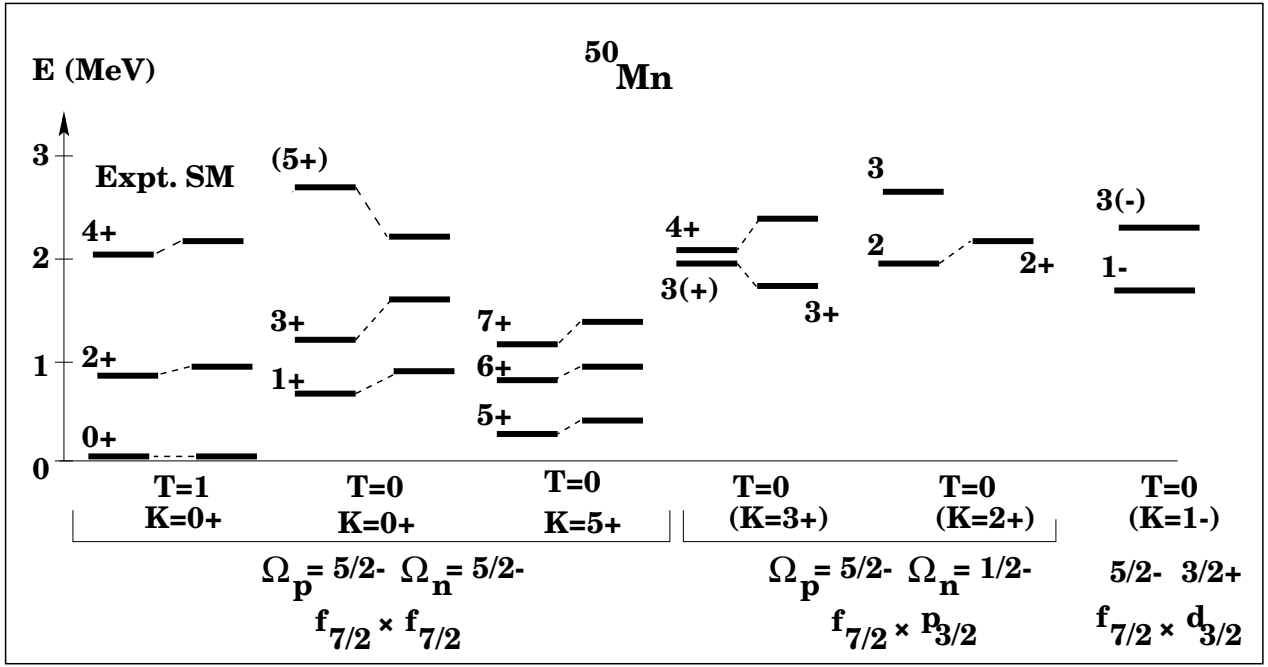}}
\vspace{1cm}
\caption{\label{theo} Comparison of the experimental (Expt.) 
 low-spin level scheme of $^{50}$Mn to the shell model results (SM) using  
 FPD6 residual interaction. The assigned K quantum number and corresponding 
 quantum numbers of the odd proton and odd neutron are shown. The values 
 given in parenthesis are based only on the results of the SM calculations.}
\end{figure*}

As an example the calculated
excitation energies with FPD6 interaction for the positive 
parity levels with spin quantum numbers $J=0-7$ below 3~MeV are
compared to the data in Fig.\,\ref{theo}. The comparison shows that the 
calculations lead  to almost perfect agreement with experiment.
Furthermore we collect SM predictions for B(E2) and B(M1) values in 
Table \ref{shellmodel} to compare with rotational model results.  

Accordingly to the Nilsson model 
last odd proton and odd neutron in  $^{50}$Mn occupy the Nilsson deformed 
$[312]5/2^-$ orbital which transforms to the spherical $f_{7/2}$ orbital 
at the limit of zero deformation. Then the low-lying states in $^{50}$Mn 
should form the lower parts of $K^\pi=0^+$,T=1 (even spins), 
$K^\pi=0^+$,T=0 (odd spins) and $K^\pi=5^+$,T=0 bands 
( see Fig.\ref{theo}). Supposing the deformation parameter $\beta$ to be 
 0.25 we have calculated B(E2) values for $K^\pi=0^+$ and 
$K^\pi=5^+$ intraband transitions (see Table \ref{shellmodel}). One can see 
that SM results are well matched by the geometrical model indicating 
band structures in $^{50}$Mn. Furthermore it follows from the Nilsson scheme that
promoting one proton or one neutron from the $[312]5/2^-$ $(f_{7/2})$ orbital 
to the closely lying $[312]1/2^-$ $(p_{3/2})$ orbital one can construct 
$K^\pi=3^+$ and $K^\pi=2^+$ bands in $^{50}$Mn. From the SM we can identify 
the $3^+_2,T=0$ (most probably it corresponds to the observed $J=3$,T=0 
level at 1798 keV) and  $4^+_2,T=0$ (it corresponds to the observed 
$4^+$,T=0 level at 1917 keV) 
states as a possible members of the $K^\pi=3^+$ band taking into account that 
they are connected by very strong E2 transition (see Table \ref{shellmodel}). 
This B(E2) value can be reproduced in geometrical model supposing that 
$K^\pi=3^+$ band is strongly deformed ($\beta \approx 0.3$). 
However such a transition was not observed in the present experiment due to the small spacing between the levels at 1798 keV and 1917 keV but there is an  
experimental indication that both states does not decay strongly to the 
states from other bands. Therefore it would be very interesting to find an 
experimental support of this theoretical hypothesis of stronger deformed 
K=3 band. 

Another interesting observation which follows from the comparison of 
theoretical and experimental data is that there are strong isovector M1 
transitions in $^{50}$Mn which have non-collective quasideuteron nature 
and could be described in frames of Nilsson model too \cite{Lis00}. 
Accordingly to the quasideuteron 
picture \cite{Lis99} (see also \cite{Zamick}) one should expect very 
large summed M1 $0^+ \rightarrow 1^+$ transition strength ( 18 $\mu_N^2$) 
for odd-odd N=Z nuclei in $f_{7/2}$ orbital. The main part of 
this strength is predicted by Nilsson model \cite{Lis00} and shell model 
to be distributed among three $0^+_1 \rightarrow 1^+_i$ transitions in 
$^{50}$Mn nucleus. From the measured decay intensity ratio for 
$2^+_1,T=1$ state in $^{50}$Mn and B(E2;$2^+_1\rightarrow 0^+_1$) value 
in the T=1 isospin partner nucleus $^{50}$Cr we have estimated ( see for 
details \cite{Schmidt})  B(M1;$2^+_1\rightarrow 1^+_1$) value in 
$^{50}$Mn which is given in Table \ref{shellmodel}. Supposing that 
theoretical and experimental ratios of B(M1;$0^+_1\rightarrow 1^+_1$) and 
B(M1;$2^+_1\rightarrow 1^+_1$) values are similar, one can actually estimate 
also that a large part of the total quasideuteron M1 
$0^+_1\rightarrow 1^+$ transition strength (up to $\sim$4.5$\mu_N^2$) could be
distributed to the $0^+_1 \rightarrow 1^+_1$ transition.

Furthermore it follows from SM calculations for 
$^{46}$V \cite{Fries98} and $^{50}$Mn  that ratios of 
B(E2;$J+2,K=0 \rightarrow J,K=0$) and B(E2;$2^+_1\rightarrow 0^+_1$) values 
are very similar for both FPD6 and KB3 interactions. Applying to this 
regularity (i.e. assuming that it is true for the experimental values) and 
using measured intensity ratios we can estimate M1 strengths
for some other transitions (see Table \ref{shellmodel}).

This estimation clearly shows that $\Delta$K=0 $\Delta$T=1 M1 transitions 
are enhanced  while other $\Delta$T=1 M1 transitions are indicated by the 
SM and experimental results to be retarded due to the K quantum number 
selection rule ($\Delta$K $>$ 1 M1 transitions are forbidden). The strong M1 
transitions can be interpreted as an consequence of considerable 
contributions of quasideuteron configurations to the low-spin K=0 states in 
$^{50}$Mn.     

To summarize, the observations for  $^{50}$Mn nucleus were compared to large 
scale shell model calculations for the positive parity states. An excellent 
agreement between theory and experiment was noted. The collective rotational 
properties of the low-lying states in $^{50}$Mn were established basing on 
the B(E2) values from shell model calculations. Strong enhancement of 
$\Delta$T=1 M1 transitions caused by the quasideuteron configurations is found
to take place for $\Delta$K=0 case. 

\section*{Acknowledgments}

We thank A.~Gelberg, J.~Eberth, K.~Jessen, R.~V.~Jolos, H.~Klein, 
V.~Werner for valuable discussions. 
This work was partly supported by the DFG under Contracts no. 
Br 799/10-2, 799/9-3, Pi 393/1-1 and by the US DOE under Contract No.
 DE-FG02-91ER-40609.

\begin{table}
\caption{\label{shellmodel} Comparison of shell model predictions ( columns 
KB3 and FPD6) for $\gamma$ transition strengths with collective rotational 
model ( column Coll.) results for $^{50}$Mn. Estimated B(M1) values following 
the procedure described in the text are given in column Est. The values of 
 K quantum number and deformation parameter $\beta$ used for the rotational 
model calculations are given. }
\begin{center}
\begin{tabular}{|ccccccc|}
\hline
    & \multicolumn{3}{c}{B(E2;$J_i\rightarrow J_f$), ($e^2$fm$^4$)} 
    & \multicolumn{3}{c|}{B(M1;$J_i\rightarrow J_f$), ($\mu_N^2 $) } \\
\cline{2-7} 
$(J_i,T_i)\rightarrow (J_f,T_f) $ & KB3 & FPD6 & Coll.& KB3 & FPD6 & Est.\\
\hline
\multicolumn{7}{|c|}{ K=0 band $\beta=0.25$} \\
\hline
$(0_1^+,0) \rightarrow (1_1^+,1)$ &  &  &  & 2.90 & 1.49 & 1.50   \\ 
$(2_1^+,1) \rightarrow (1_1^+,0)$ & 0.05 & 0.02 & 0 & 1.94  &1.29 & 1.00  \\ 
$(2_1^+,1) \rightarrow (0_1^+,1)$ & 220 & 275 & 202 &    &    &  \\
$(3_1^+,0) \rightarrow (2_1^+,1) $& 0.0 & 0.001 & 0 & 3.73 & 1.92 & 1.25 \\ 
$(3_1^+,0) \rightarrow (1_1^+,0) $& 272 & 350 & 260 &      &      &  \\
$(4_1^+,1) \rightarrow (3_1^+,0) $& 0.2 & 0.07 & 0 & 2.71 & 1.99 & 1.34  \\ 
$(4_1^+,1) \rightarrow (2_1^+,1) $& 298 & 385 & 289&  &  & \\ 
$(5_2^+,0) \rightarrow (4_1^+,1)   $& 0.4  & 1.4  & 0 & 3.46 & 2.11 & 0.77 \\ 
$(5_2^+,0) \rightarrow (3_1^+,0) $& 227 & 303 & 306 & & & \\
\hline 
\multicolumn{7}{|c|}{ K=5 band $\beta=0.25$} \\
\hline
$(6_1^+,0) \rightarrow (5_1^+,0)$ & 293  & 373  & 305 &1$\cdot10^{-4}$ & 0.003 &  \\
$(7_1^+,0) \rightarrow (6_1^+,0)$ & 285  & 385 & 361 & 3$\cdot10^{-4}$ & 0.01 &
 \\ 
$(7_1^+,0) \rightarrow (5_1^+,0)$ & 48  & 56  & 49 & 0 & 0 &  \\
\hline
\multicolumn{7}{|c|}{  K=3 band $\beta=0.3$} \\
\hline 
$(4^+_2,0) \rightarrow (3_2^+,0) $& 512  & 544  & 526 & 1$\cdot10^{-5}$ & 0.015 & \\ 
\hline 
\end{tabular}
\end{center}
\end{table}  

\end{document}